\documentclass[preprint,12pt]{elsarticle}
\usepackage[utf8]{inputenc}
\usepackage{hyperref}
\usepackage{graphicx}
\usepackage{caption}
\usepackage{subcaption}
\usepackage{tikz}

\journal{}

\begin{document}
\newcommand{\nbseedpapers}[0]{$30$}
\newcommand{\nbquerypapers}[0]{$53$}
\newcommand{\nbsnowpapers}[0]{$70$}
\newcommand{\nbpapers}[0]{$25$}

\begin{frontmatter}

\title{Applications of statistical causal inference in software engineering.}

\author[inst1]{Julien Siebert}

\address[inst1]{Data Science Department, Fraunhofer Institute for Experimental Software Engineering (IESE),\\
            Fraunhofer Platz 1, 
            67663 Kaiserslautern,\\
            Rhineland-Palatinate,
            Germany}

\begin{abstract}
\textbf{Context:} The aim of statistical causal inference (SCI) methods is to estimate causal effects from observational data (i.e., when randomized controlled trials are not possible). In this context, Pearl's framework based on causal graphical models is an approach that has recently gained popularity and allows for explicit reasoning about issues related to spurious correlations. \\
\textbf{Objectives:} Our primary goal is to understand to which extend and how Pearl's graphical framework is applied in software engineering (SE).  \\
\textbf{Methods:} We performed a systematic mapping study and analysed a total of \nbpapers{} papers published between 2010 and 2022.\\
\textbf{Results:} Our results show that the application of Pearl's SCI framework in SE is relatively recent and that the corresponding research community is fragmented. Most of the selected papers focus on software quality analysis. There is no clear and widespread community of practice (yet) on how to implement and evaluate SCI in SE.\\
\textbf{Conclusions:} To the best of our knowledge this is the first time such a mapping study is done. We believe that SE practitioners might benefit from such a work, as it both provides an overview of the work and people involved in the application of causal inference methods, but also outlines the potential and limitations of such approaches. 
\end{abstract}




\end{frontmatter}

\section{Introduction}
Empirical software engineering aims at studying and evaluating how people engineer software with the help of empirical research methods. This paper focuses on the application of one type of empirical methods, namely statistical causal inference (SCI, see section \ref{sec:def}). Such methods have their roots in a number of applied fields (from AI to econometrics) and aim to provide a framework for making valid inferences about causal effects based on interventional or observational data. More specifically, we focus on SCI methods that use graphical models as developed by Pearl and colleagues \cite{pearl2016causal,pearl2019book}. This framework has been shown to be equivalent of the potential-outcomes framework (also called the Neyman-Rubin Causal Model \cite{yao2021survey}) but enriches it by making use of an explicit causal structure called a graphical causal model. Making assumptions about causal effects explicit through a graphical structure has several advantages. First, it helps determine whether causal effects can be estimated and how they might be estimated (see section \ref{sec:def}). It helps to find out whether an analysis might suffer from problems such as confounder or collider bias \cite{elwert_collider_2014,cinelli2022crashcourse,huenermund_doubleml_2021}. Finally, the hypotheses represented in the graphical causal model are amenable to falsification because they imply a certain set of independencies that can be tested in the data \cite{pearl2019book}.

In short, causal graphical models can be seen as a basic building block of causal inference. However, they are relatively new compared to the potential outcomes framework, and methods for estimating causal effects from observational data that do not explicitly rely on a causal graphical model (such as matching, difference-in-differences, instrumental variables, or regression discontinuity, see \cite{huntington2021effect,cunningham_mixtape_2021}) existed long before the theory behind causal graphical models was invented. Furthermore, depending on the system under study, it may be difficult to obtain a sufficiently detailed causal graphical model: some influencing factors may not be directly measurable and only available through proxy data, or may even be unknown. A typical example in SE would be aspects such as "developer experience" or "effort and time to develop a given feature" \cite{trendowicz_techdebt_2021}. 

The goal of this paper is to understand whether, how and by whom Pearl's framework for causal inference is applied in software engineering (SE) (see section \ref{sec:methodogy} for the detailed research questions). We hope to provide an overview of the field that helps evaluating the potential and limitations of these SCI methods for SE. For that, we performed a literature survey following the guidelines from Kitchenham and Charters \cite{kitchenham_guidelines_2007}, and Wohlin \cite{wohlin2014guidelines}. We analysed a total of \nbpapers{} papers. To the best of our knowledge, this is the first time such a mapping study is done.

This paper is structured as follows. Section \ref{sec:def} provides a short introduction to SCI methodology. Section \ref{sec:related_work} provides a short analysis of the related work. Section \ref{sec:methodogy} presents our search methodology, section \ref{sec:analysis} our analysis and section \ref{sec:answers} our results. We discuss and conclude our work in both sections \ref{sec:discussion} and \ref{sec:conclusion}.

\section{Definitions}\label{sec:def}
\subsection{Statistical Causal Inference (SCI)}
Statistical causal inference focuses on estimating actual causal effects of an action (a treatment $T$) on a given observed system (an outcome $Y$) from data \cite{pearl2016causal,hernan2020whatif,yao2021survey}. In general, the gold standard for estimating a causal effect is to perform randomized controlled trials in which the potential confounders\footnote{A confounder is a variable that both have an impact on the treatment and the outcome of interest. See figure \ref{fig:confounder}.} are controlled for, through, e.g., randomization, so that their effects are balanced between the experimental and the control groups and that only the treatment differs between groups. However, when this is not feasible (e.g., for practicable or ethical reasons), causal effects have to be estimated from observational data (i.e., data that was not generated during randomized controlled trials). The fundamental problem with observational data is that we cannot have the same subject being treated and not being treated at the same time (following Pearl's notation and assuming a binary treatment, we cannot observe $P(Y|\mathrm{do}(T=0))$ and $P(Y|\mathrm{do}(T=1))$ at the same time) and the effect has to be estimated from available data (i.e., from $P(Y|T=0)$ and $P(Y|T=1)$). The challenge with observational data, assuming that all relevant variables have been captured, is that it can contain correlations induced by confounding factors that can bias the estimation of causal effects (see figure \ref{fig:confounder}). For this reason, researchers in the field of statistical causal inference have developed methods to reduce the impact of confounding factors and separate spurious correlations from causal effects \cite{huntington2021effect,cunningham_mixtape_2021}. One of such approach is the use of graphical causal models together with the application of the do-calculus by Pearl et al. \cite{pearl2016causal,pearl2019book}.

\subsection{Graphical causal models}\label{ref:causal_graph}
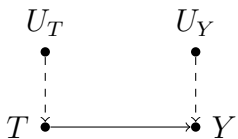
\begin{figure}[!htbp]
    \centering
        \begin{tikzpicture}
            \node (ut) at (0,1) [circle,fill,inner sep=1.25pt,label=above:$U_T$] {};
            \node (uy) at (2,1) [circle,fill,inner sep=1.25pt,label=above:$U_Y$] {};
            \node (t) at (0,0) [circle,fill,inner sep=1.25pt,label=left:$T$] {};
            \node (y) at (2,0) [circle,fill,inner sep=1.25pt,label=right:$Y$] {};
            \path[->] (t) edge (y);
            \path[->,dashed] (ut) edge (t);
            \path[->,dashed] (uy) edge (y);
        \end{tikzpicture}
    \caption{Example of a graphical causal model representation where a treatment $T$ causes an outcome $Y$. $T$ and $Y$ are endogenous variables. $U_T$ and $U_Y$ are exogenous variables (e.g., external noise).}
    \label{fig:graphical_causal_model}
\end{figure}

Graphical causal models are a way to represent hypotheses about causality in a given system. They take the form of a graph\footnote{Causal graphs are by definition directed acyclic graph and are sometimes referred to (causal) DAG.} consisting of nodes representing the relevant variables (potentially measurable) and links representing a direct causal effect between two variables. For example, if a treatment $T$ is supposed to have a causal effect on an outcome $Y$, then  both $T$ and $Y$ are represented as nodes and a link goes from $T$ and points towards $Y$. Variables can be either endogenous (they belong to the system under study) or exogenous (they are outside the system). Observed effects are usually represented by solid lines, while unobserved effects are represented by dashed lines. Figure  \ref{fig:graphical_causal_model} illustrates a simple example of a graphical causal model. Graphical models are interesting for different reasons. First, they make hypotheses about variables under study and their causal effects explicit. Second, they are used for identifying structures that lead to spurious correlations and need to be handled before estimation. Third, causal graphical models can be used as a visualization help to understand in which case other SCI methods (such as propensity score matching, difference-in-difference, instrumental variables, etc.) are applicable (see section \ref{ref:other_sci_methods}).

\subsection{Pearlian statistical causal inference methodology} 
The methodological process behind Pearl's SCI consists of three major steps: modeling, identification, estimation\footnote{Pearl and Mackenzie introduced a 9 steps process (see Figure 1 in \cite{pearl2019book}). Sharma and Kiciman proposed 4 steps in \cite{sharma2020dowhy}. Both resolve around the three steps mentioned above but introduce some more details and testing methods that are out of this paper's scope. For the sake of simplicity, we just introduce the three steps of modeling, identification and estimation.}. The first step (modeling) is to make causal assumptions explicit through the use of a graphical causal model. This causal model can either be obtained through domain expertise or extracted from data. In the latter case, the term "causal structure discovery" or "causal structure learning" is used (see  \cite{glymour2019review,vowels2022survey}).

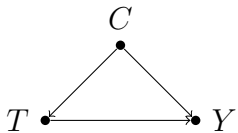
\begin{figure}[!htbp]
    \centering
        \begin{tikzpicture}
            \node (c) at (1,1) [circle,fill,inner sep=1.25pt,label=above:$C$] {};
            \node (t) at (0,0) [circle,fill,inner sep=1.25pt,label=left:$T$] {};
            \node (y) at (2,0) [circle,fill,inner sep=1.25pt,label=right:$Y$] {};
            \path[->] (t) edge (y);
            \path[->] (c) edge (t);
            \path[->] (c) edge (y);
        \end{tikzpicture}
    \caption{Graphical representation of a backdoor path where $C$ is a confounder that infuences both treatment $T$ and outcome $Y$. The backdoor path is $T \leftarrow C \rightarrow Y$. Estimating the effect of $T$ on $Y$ requires to "close" the backdoor. In the present case it can be done by adjusting for $C$. Following Pearl's notation and assuming that $T$ is binary, the causal effect of $T$ on $Y$ is written  $P(Y=y|\mathrm{do}(T=0))-P(Y=y|\mathrm{do}(T=1))$. The do operator represents the fact that we intervene on the system by setting $T$ to a specific value. Adjusting for $C$ implies $P(Y=y|\mathrm{do}(T=0))=\Sigma_c P(Y=y|T=0,C=c)P(C=c)$ and $P(Y=y|\mathrm{do}(T=1))=\Sigma_c P(Y=y|T=1,C=c)P(C=c)$. The absence of do operators on the right-hand side of the equations tells us that we can estimate the causal effect from observational data alone: $P(Y=y|\mathrm{do}(T=0))-P(Y=y|\mathrm{do}(T=1))=\Sigma_c P(Y=y|T=0,C=c)P(C=c)-\Sigma_c P(Y=y|T=1,C=c)P(C=c)$.}
    \label{fig:confounder}
\end{figure}

The second step (identification) is to find out whether a given causal effect can be estimated using the available data. 
The idea is to find out whether there are paths in the graph where a variable can influence both the treatment and the outcome of interest. These are called backdoor paths  and are sources of spurious correlations if left "open" (see figure \ref{fig:confounder} for a visual example of how a backdoor path can be closed). Identifying such path requires identifying structures such as confounders, mediators, or colliders to understand which variable(s) should be controlled for \cite{cinelli2022crashcourse}. Identification outputs an estimand for the causal effect under study. The main tool for causal identification is the do-calculus developed by Pearl and colleagues \cite{pearl2012docalculus}. 

The third step (estimation) is to effectively estimate the identified causal effect based on the available data and the previously gained knowledge of which variables should be adjusted for. Estimation can be done using a variety of methods, from simple linear regression to more complex approaches such as T-learner, X-learner, double machine learning or causal random forests\footnote{The interested reader may wish to refer to Chapter 7 in \cite{brady2020introduction} for an overview of estimation methods}. 

Like any other analytical method, SCI based on a causal graphical model relies on certain assumptions. First, the graph should include all relevant influencing factors (measurable or not) for the problem at hand, as unknown confounders could make the analysis unreliable (this is called the unconfoundedness assumption). Second, the statistical properties implied by the graph (such as conditional independence) should represent those that exist in reality (this is called the Markov assumption). This means that the conditional independencies implied by the graph can be tested in the available data, making the causal graph amenable to falsification. Different assumptions also need to be considered in estimation methods. For example, the stable unit treatment value assumption (SUTVA) states that the outcome of a given unit should not change due to the treatment of other units. The positivity assumption states that for a given subgroup both treated and untreated individuals should be present \cite{brady2020introduction,yao2021survey}. Finally, several tests have been proposed for refuting causal estimates, like placebo test (see \cite{sharma2020dowhy}).

\subsection{Relation to Bayesian networks and structural equation models}
The focus of this paper is on causal inference methods using graphical causal models, therefore it is necessary to sketch the distinction between these models and others like structural equations models and Bayesian networks.

The graphical models in SCI are indeed are very similar to Bayesian networks, with the constraints that an arrow represents a direct causal effect, and that no arrow implies that no direct causal effect exists between variables. The main difference lies in the inference process. While Bayesian networks allow for probabilistic inference and prediction based on correlations, causal inference requires an additional step, namely identification. As stated before, the identification step is to find out whether a given causal effect can be estimated using the available data and tells us which variable should be controlled for (see examples of good and bad controls in \cite{cinelli2022crashcourse}).

Another type of graphical models are structural equation models (SEM), which represent cause and effects as equations. They sometimes use a specific symbol $:=$ to explicitly state that a variable is the cause of another one (for example $Y:=f(X)$ means that $X$ is a cause of $Y$), and that one cannot simply reverse the equation (e.g., $Y:=f(X)$ does not mean $X:=f^{-1}(Y)$). These models provide a functional form of the effect (e.g., linear, non-linear). In comparison, graphical causal models do not provide any functional form of the effect (they are also called non-parametric). As such, structural equation models can be viewed as a subclass of graphical causal models, which in turn are themselves a subclass of Bayesian networks.

\subsection{Other methods not explicitly based on a causal graphical model}\label{ref:other_sci_methods}
\begin{figure}[!hbtp]
    \centering
        \begin{tikzpicture}
            \node (c) at (1,1) [circle,fill,inner sep=1.25pt,label=above:$C$] {};
            \node (iv) at (-2,0) [circle,fill,inner sep=1.25pt,label=above:$IV$] {};
            \node (t) at (0,0) [circle,fill,inner sep=1.25pt,label=below:$T$] {};
            \node (y) at (2,0) [circle,fill,inner sep=1.25pt,label=right:$Y$] {};
            \path[->] (t) edge (y);
            \path[->,dashed] (c) edge (t);
            \path[->,dashed] (c) edge (y);
            \path[->] (iv) edge (t);
        \end{tikzpicture}
    \caption{Typical causal graph assumed by the instrumental variables estimation method. $C$ is an unobserved confounder. $IV$ is the instrumental variable. Note that for this method to work, there must be no open backdoor from $IV$ to $Y$.}
    \label{fig:dag_iv}
\end{figure}
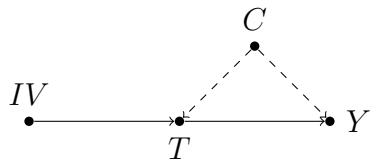

The use of causal graphical models is relatively recent (see Pearl's historical perspective in \cite{pearl2019book}), and the do-calculus can be seen as a theoretical cornerstone that allows for systematic reasoning about association and causal flows \cite{cinelli2022crashcourse}. However, for some application cases, it may be difficult to obtain a sufficiently detailed graphical causal model for the do-calculus to be effective in providing a causal effect estimand. There are many causal inference methods that do not explicitly rely on the modelling, identification, estimation phases mentioned above. Some methods either compare treated and untreated individuals who are assumed to be similar except for treatment (such as matching or regression discontinuity), or modify their variables to make them more similar (like re-weighting methods). 
Other methods, such as event studies, difference-in-differences or synthetic control, attempt to model the counterfactual (i.e. what would have been the behaviour of the treated group if no treatment had been given) and measure the difference between reality and the assumed counterfactual.
Finally, some methods, such as instrumental variables, work well for specific causal structures (see figure \ref{fig:dag_iv}). In all cases, a causal graphical model helps understand when the method applies and whether the underlying assumption of a method still holds for a given application case (see how recent textbooks on causal inference make use of the causal graph to explain the assumptions behind estimations methods \cite{huntington2021effect,hernan2020whatif,cunningham_mixtape_2021}).

\section{Related Work}\label{sec:related_work}
\subsection{Generic causal inference surveys}
While technical SCI surveys typically focus on technical aspects (such as issues related to the type of data, like time-series \cite{moraffah_2021_timeseries} or geospatial data \cite{akbari2023spatial}) and are usually domain agnostic, they can provide insight into application domains and their corresponding use cases. For example, \cite{yao2021survey, moraffah_2021_timeseries} discusses the potential of SCI methods for specific application domains along with related research and lists benchmark datasets. Software engineering is not treated directly as an application domain in these papers. Furthermore only two benchmark data sets are related to software engineering and consist of social media usage (Blogger and Flickr). In both reviews \cite{yao2021survey,kaddour_2022_causalml}, the authors discuss applications of SCI that are relevant to the engineering of data-driven software components (SE4ML/SE4AI): fairness and explainability, or for improving machine learning in general (see also \cite{cui_2022_stable,plecko2022causal}). 

\subsection{Reviews of SCI in other domains}
Domain-specific surveys can provide information about the benefits and applicability of SCI methods within a particular domain.
In \cite{schuessler_2019_survey}, the authors discuss the use of graphical causal models for survey research and how these tools can shed light on issues such as selection bias and help to question current practice. In \cite{natcom2021medicine,castro2020matters,richens2020diagnosis}, the authors discuss the application of SCI in the medical field. While they also mention the mitigation of selection bias as one aspect where graphical models can help, they also mention the mitigation of spurious correlation as a major advantage of relying on SCI methods over correlational ones (this is also mentioned in other domains, see \cite{ohlsson2020applying,burton2023earthquake}). A canonical example of a spurious correlation reported in this field comes from experiments attempting to predict the likelihood of death from pneumonia using historical medical records. Counterintuitively, the analyses found that asthma lowered the risk of death, even though it is known to be a severe condition in people with pneumonia. This misleading effect occurred because the patients with asthma in the training data set were treated more drastically than other patients. This level of care is a confounding factor that, if not taken into account, could lead to the wrong decision \cite{richens2020diagnosis,ferrari_2022_covid}. These reviews mention the other advantages, such as dealing with data drift or explainability \cite{natcom2021medicine,castro2020matters}. The arguments provided are consistent with those mentioned in the articles in the previous section.
Finally, these domain-specific reviews highlight the challenges in applying SCI. Specifically for those relating to graphical causal models, the main problem is to obtain such a graph. Another challenge mentioned is that SCI is not part of the traditional curricula in these fields.

\subsection{Causal inference in Software Engineering}
It is not the first time that the application of a given analysis method in SE is reviewed. The application of Bayesian networks in SE is relatively well established. For example, de Sousa et al. \cite{de_sousa_20year_2022} performed a mapping study in which they analysed 109 publications (from 1999 to 2018) focused on the application of Bayesian networks for software project management. Another mapping study was done by Misirli and Bener \cite{misirli_mapping_2014} with a focus on software quality prediction. Other systematic literature reviews and comparative analyses have also been performed \cite{del_aguila_bayesian_2016,tosun_systematic_2017,mendes2019using}. However, none of these papers mention SCI as defined in section \ref{sec:def}. 

The work by Wong \cite{wong2020computational} focuses on challenges related to developing software for causal inference (which the author called "computational causal inference"). The focus is symmetrical to that of the current paper, in that Wong focuses on the application of SE for SCI, whereas we focus on the application of SCI for SE.

To the best of our knowledge, there is no similar review like ours. The closest we could find is an  extensive related work section in \cite{clark_testing_2022} that refers to several papers using SCI in software testing but does not attempt to organize nor analyse them.


\section{Methodology}\label{sec:methodogy}
\subsection{Frame}
\subsubsection{Research questions}
Our primary goal in this paper was to identify applications of SCI methods (as defined in section \ref{sec:def}) in SE and to understand how these methods are used. We refined this goal into specific questions listed in table \ref{tab:research-questions}.We particularly focused on an analysis of the existing literature to answer our research questions. Table \ref{tab:information_extracted} lists the information, we extracted from the papers. Table \ref{tab:exclusion-criteria} provides the detailed exclusion criteria. Note that we did not limit our literature search by date. However, as the do-calculus was only invented around 1995 \cite{pearl2012docalculus}, we did not expect to find any relevant paper before 2005 and would not have included any paper prior to 1995.

\begin{table}[!hbtp]
    \footnotesize
    \centering
    \begin{tabular}{|c|p{0.8\textwidth}|}
        \hline
        \textbf{Id} & \textbf{Research Question} \\
		\hline
		\textbf{RQ1} & What are the current use cases of SCI in SE? \\
        \hline
        \textbf{RQ2} & How is the research community using SCI in SE structured?\\
		\hline
        \textbf{RQ3} & Which causal inference methods were used for each use case? \\
        \hline
        \textbf{RQ4} & Which data sets and libraries were used? \\
        \hline
        \textbf{RQ5} & How authors evaluate their results? \\
        \hline
    \end{tabular}
    \caption{Research Questions.}
    \label{tab:research-questions}
\end{table}

\subsubsection{Exclusion criteria}
\begin{table}[!htbp]
    \footnotesize
    \centering
    \begin{tabular}{|c|p{0.8\textwidth}|}
        \hline
        \textbf{Id} & \textbf{Exclusion Criteria} \\
        \hline
        \textbf{EC1} & The document contains no reference to SCI (as defined in section \ref{sec:def}). \\
        \hline
        \textbf{EC2} & The application is outside the scope of SE (as defined by the SWEBOK\footnote{\url{http://swebokwiki.org/Main_Page}}). \\
        \hline
        \textbf{EC3} & The text presents a software library dedicated to causal inference. \\
        \hline
        \textbf{EC4} & There is no access to the article. \\
        \hline
        \textbf{EC5} & The article is not written in English. \\
        \hline
        \textbf{EC6} & The paper is a thesis. \\
        \hline
        \textbf{EC7} & The paper is not a primary study. \\
        \hline
        \textbf{EC8} & The paper is a duplicate or a companion paper. \\
        \hline
    \end{tabular}
    \caption{Exclusion criteria.}
    \label{tab:exclusion-criteria}
\end{table}

\subsubsection{Extracted information}
To answer our research questions, we first extracted demographics like authors name, affiliation and type of publication. This helps learning about the structure of the research community using SCI in SE (RQ2). We extracted information about the SE use cases (RQ1). To answer RQ3 and RQ4, we extracted information about the SCI methods, the data sets and the libraries used. Finally, we extracted information about the evaluation of the results (RQ5). Table \ref{tab:information_extracted} provides an overview of the information collected. 

\begin{table}[!htbp]
    \footnotesize
    \centering
    \begin{tabular}{|p{0.3\textwidth}|p{0.6\textwidth}|}
        \hline
        \textbf{Information extracted} & \textbf{Description} \\
        \hline
        Authors demographics & authors names and affiliation. \\
        \hline
        Type of paper & journal, conference paper, report, preprint, etc. \\
        \hline
        SE use case & e.g., fault detection, effort estimation, etc. \\
	    \hline
        Causal analysis task done and corresponding method name & Causal discovery, identification, estimation.\\
        \hline 
        Data sets & name and URL.\\
        \hline 
        Libraries & name and URL.\\
        \hline
        Evaluation methods and metrics & Corresponding name and, when available, formal description. \\
        \hline
    \end{tabular}
    \caption{Information extracted.}
    \label{tab:information_extracted}
\end{table}

\subsection{Search and selection process}
There are mainly two difficulties here. First, SE is a large field and filtering relevant papers (i.e., use cases that belong or do not to SE) can be challenging. For this, we relied upon categories of activities defined in the SWEBOK\footnote{\url{http://swebokwiki.org/Main_Page}}. Second, SCI methods are relatively new and, in general, may be less well known than statistical data analysis methods. In addition, graphical causal models are close relatives of Bayesian networks and methods such as the do-calculus could have been applied without direct reference to the generic term "causal inference". 

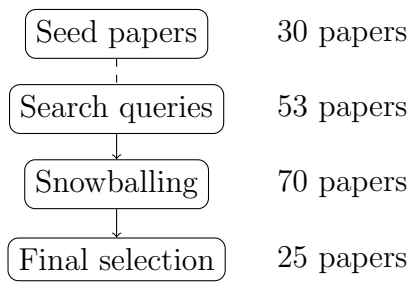
\begin{figure}[!htbp]
    \centering
        \begin{tikzpicture}
            \node[draw,rounded corners](seed) at (0,0) {Seed papers};
            \node[draw,rounded corners](query) at (0,-1) {Search queries};
            \node[draw,rounded corners](snow) at (0,-2) {Snowballing};
            \node[draw,rounded corners](filter) at (0,-3) {Final selection};
            \node[](seed_p) at (3,0) {\nbseedpapers{} papers};
            \node[](query_p) at (3,-1) {\nbquerypapers{} papers};
            \node[](snow_p) at (3,-2) {\nbsnowpapers{} papers};
            \node[](filter_p) at (3,-3) {\nbpapers{} papers};
            \path[-,dashed] (seed) edge (query);
            \path[->] (query) edge (snow);
            \path[->] (snow) edge (filter);
        \end{tikzpicture}
    \caption{Search process overview.}
    \label{fig:search_process}
\end{figure}

The search process and selection followed four main steps: 1) establishing a set of seed documents, 2) consolidating a search query, 3) extending the query results with a snowballing phase, and 4) selecting the final set of relevant papers (see Figure \ref{fig:search_process}). For the first three steps, the selection was done based on title and abstract screening. In the last step, the selection was done after full-read.

\subsubsection{Seed documents}
We started our search with a relatively simple query in Scopus\footnote{\url{https://www.scopus.com}} completed with a snowballing phase (both backward and forward) in order to find a first set of seed papers. The following query: \texttt{TITLE-ABS-KEY((\{causal inference\} OR \{causal discovery\} OR \{structure learning\}) AND (\{SE\}))} was performed on 26.04.2022 and resulted in 42 documents among which 11 were judged relevant (after title and abstract screening). 

A forward and backward snowballing pass was done on these first 11 documents. This led to 19 new relevant papers for a total of \nbseedpapers{} seed papers. 

These \nbseedpapers{} articles were used, on the one hand, as seed articles (i.e., to test whether subsequent search queries would include these articles) and, on the other hand, to gather information about how researchers in the field of SE describe SCI and find interesting search terms that helped us find more relevant articles.

\subsubsection{Search query}
In the next phase, we created a search query that would include our seed papers and return new potentiall relevant papers. We actually created two search queries (see figure \ref{fig:search_query}).  The first query looks for papers both mentioning SE and SCI. The second query looks for papers citing the work of Judea Pearl.

\paragraph{First query}\label{par:first_query}
As stated above, the first query is structured in two parts. Each part contains relevant keywords for each topic of interest and all keywords are separated by \texttt{OR}. Both parts are separated by \texttt{AND}.

The search for relevant query terms for both parts was done as follows. First, single SE terms were tested in conjunction with "causal inference". The SE terms were chosen both from the SWEBOK and from the seed papers. Then, we checked the number of papers returned by Scopus as well as their relevance to our research questions. The assumption was that the keywords return a higher ratio of relevant papers and that the constraint \texttt{AND "causal inference"} is generic enough to keep most of the relevant papers but focused enough to exclude many irrelevant ones.

A similar approach was taken for determining relevant keywords related to SCI. The chosen terms were taken from reviews and textbooks \cite{elwert2013graphical,pearl2016causal,brady2020introduction,hernan2020whatif,yao2021survey} and consolidated with a short overview of the vocabulary used in the field of SCI\footnote{Here the author performed a search in Scopus using the query 
\texttt{(TITLE("causal inference")) AND ("graph*")}, extracted the main authors (i.e., authors having more than 5 documents within this query), extracted related and relevant topics using SciVal \texttt{https://scival.com/} (Topic T.12463: "Chain Graph; Artificial Intelligence; Conditional Independence", Topic T.1910: "Observational Studies; Causal Inference; Propensity Score", and Topic T.22578: "Principal Stratification; Causal Effect; Outcome"), and finally extracted key phrases from these topics.}. Single causal inference terms were tested in conjunction with "Software Engineering" (i.e., adding \texttt{AND "Software Engineering"} to the query).

Table \ref{tab:prep-query} gives the details of the queries tested and the related hits and relevant papers. The final query was built using the more encompassing terms (i.e., terms that provided more hits and more relevant papers but also included relevant papers from other queries). The final query is shown in figure \ref{fig:search_query}.

\paragraph{Second query}\label{par:second_query}
The second query targeted papers citing Pearl's work and mentioning "Software Engineering" and "causal inference" in their text. Scopus allows for searching through all the papers citing a given authors by using the author's scopus ID ("Pearl, Judea" 7101604154). The query is also shown in Figure \ref{fig:search_query}.

\begin{figure}[!htbp]
    \textbf{First query:}\\
    \begin{center}
        \texttt{
        TITLE-ABS-KEY(
        ("software engineering"  OR  "software test*"  OR  "software quality"  OR  "fault local*"  OR  "AIOps") 
        AND
        ("causal inference"  OR  "propensity score"  OR  "counterfact*"  OR  "inverse probability"  OR  "probability weigh*" ))  
        }
    \end{center}
    ~\\
    \textbf{Second query:}\\
    \begin{center}
        \texttt{
        "Pearl, Judea" 7101604154 \\
        Refined to [( "causal inference"  AND  "software engineering" ) ]
        }
    \end{center}
    \caption{Search queries.}
    \label{fig:search_query}
\end{figure}

\subsubsection{Snowballing phase}
These two queries returned 86 and 119 papers (as of 2022-11-15), out of it, a total of  \nbquerypapers{} distinct papers were judged relevant based on title and abstract screening. These papers were used as seed for snowballing. Backward snowballing was done using Scopus directly. Forward snowballing was done using Google Scholar. Relevance of the papers was judged based upon title and abstract screening. We performed two snowballing phases. First, backward and forward snowballing was done on the first set of \nbquerypapers{} papers. This led to 29 new papers (backward: 8, forward: 21). Second, backward and forward snowballing was done on this novel set of 29 papers. The snowballing phases allowed us to collect 95 potentially relevant papers.

After reading the papers completely, we excluded 25 papers. 18 papers were judged out of scope because they were not using SCI as defined in section \ref{sec:def} (EC1). One paper was not written in English (EC5), one paper was describing a software library (EC3), and there were two papers that we didn't have access to (EC4). One paper was a secondary study (a review) (EC7), one paper was a duplicate and one was a companion paper (EC8).  This led to a total of \nbsnowpapers{} papers that we further analysed.

\subsubsection{Papers categorization and final selection}
During the information extraction phase, we noticed that the articles fell into four distinct categories. 
The first category (C1), contains papers focusing only on the first step of causal inference (modeling) and do not mention the next two steps (identification and estimation). 22 papers are present in this category (cf. Table \ref{tab:category_c1}). Note that in this category, many papers use the term causal analysis or causal inference when talking about causal structure discovery.

The second category (C2), contains papers focusing on counterfactual analysis (also called counterexamples or causality checking) but not doing causal inference as defined in section \ref{sec:def}. Although, they make use of structural equation models (SEM) and the notion of counterfactual as defined by Halpern and Pearl \cite{halpern2005causes}, the papers in this category did not use SCI methods (at least none of the paper, we reviewed was mentioning either the different phases of causal inference or specific techniques like do-calculus for identification or specific estimation methods). 16 papers were classified as C2 (cf. Table \ref{tab:category_c2}).

The third category (C3), contains papers doing causal inference without relying upon a graphical model. These papers use techniques (like propensity score matching or double machine learning) to handle confounding bias. Seven papers have been classified as C3 (cf. Table \ref{tab:category_c3}).

The last category (C4), contains papers doing all three steps of causal inference: modeling, identification, and estimation. \nbpapers{} papers were classified as C4 (cf. Table \ref{tab:category_c4}).

With regard to our research questions, we decided to exclude categories C1, C2, and C3 from further analysis. The reason is twofold. First, our paper focuses on the application of Pearl's framework for causal inference and the papers in C1, C2, and C3 categories do not entirely fulfill this requirement. Second, our search was primarily designed to find papers from the C4 category, and therefore, we suspect that the papers from categories C1, C2, and C3 are only a subset of what we could have found if we would have directly targeted either causal structure discovery, counterfactual analysis, or causal inference techniques not relying upon a causal graphical model.

\section{Analysis}\label{sec:analysis}
\subsection{Demographics and bibliometrics}\label{sec:demographics}
\begin{figure}[!htbp]
    \centering
    \includegraphics[width=0.6\textwidth]{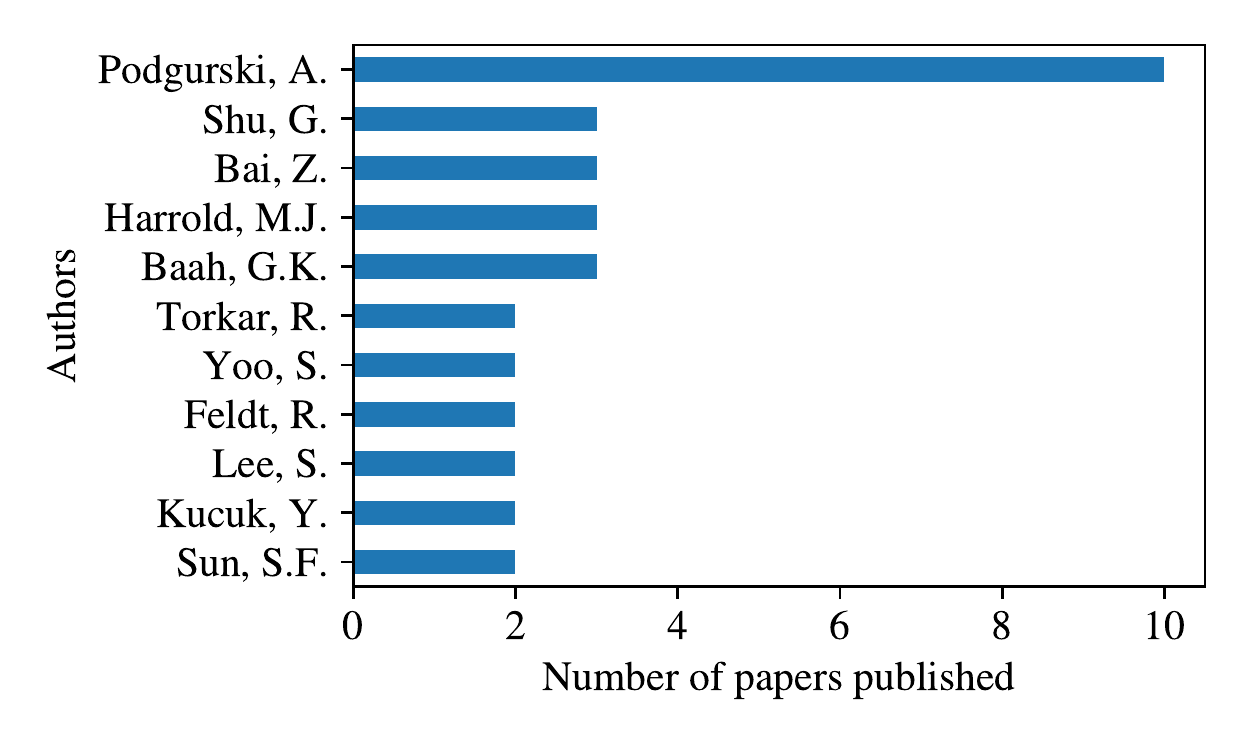}
    \caption{Number of publications per authors. Only authors being listed in more than one paper are displayed.}
    \label{fig:nb_publications_authors}
\end{figure}

The \nbpapers{} papers classified in C4 have been written by a total of 60 distinct authors. Out of theses 60 persons, only 11 are listed as authors in more than 2 papers (see figure \ref{fig:nb_publications_authors}). From the coauthorship network, we obtained 13 distinct coauthors groups (i.e., authors having published together). The largest coauthors group (in terms of papers collected) accounts for 10 papers \cite{baah_causal_2010,baah_matching_2011,baah_mitigating_2011,bai_importance_2015,bai_numfl_2015,bai_causal_2017,kucuk_improving_2021,podgurski_counterfault_2020,shu_mfl_2013,sun_properties_2016} and is composed of the following persons:
Baah, G.K. and Harrold, M.J. both from Georgia Institute of Technology Atlanta, GA; Bai, Z., Cao, F, Kucuk, Y., Shu, G., Sun, B., Sun, S-F., and Podgurski, A. from Case Western Reserve University, Cleveland, OH, USA; and Henderson, T.A.D. from Google Inc. Mountain View, CA, USA.

\begin{figure}[!hbtp]
    \centering
    \begin{subfigure}[t]{0.475\textwidth}
        \centering
        \includegraphics[width=\textwidth]{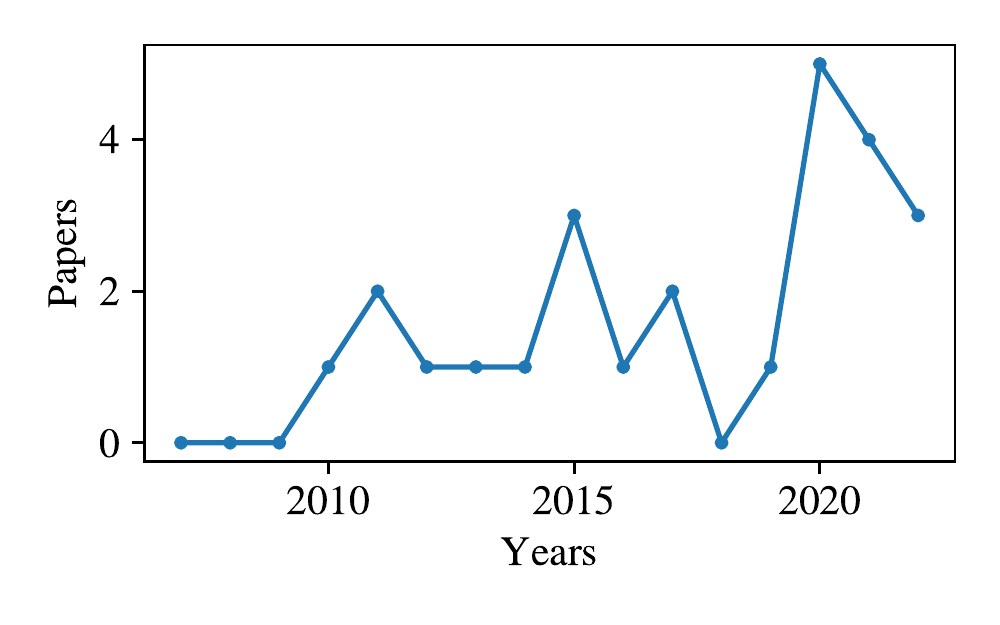}
        \caption{Evolution of the number of publications over time.}
        \label{fig:papers_per_year}
     \end{subfigure}
     \hfill
     \begin{subfigure}[t]{0.475\textwidth}
        \centering
        \includegraphics[width=\textwidth]{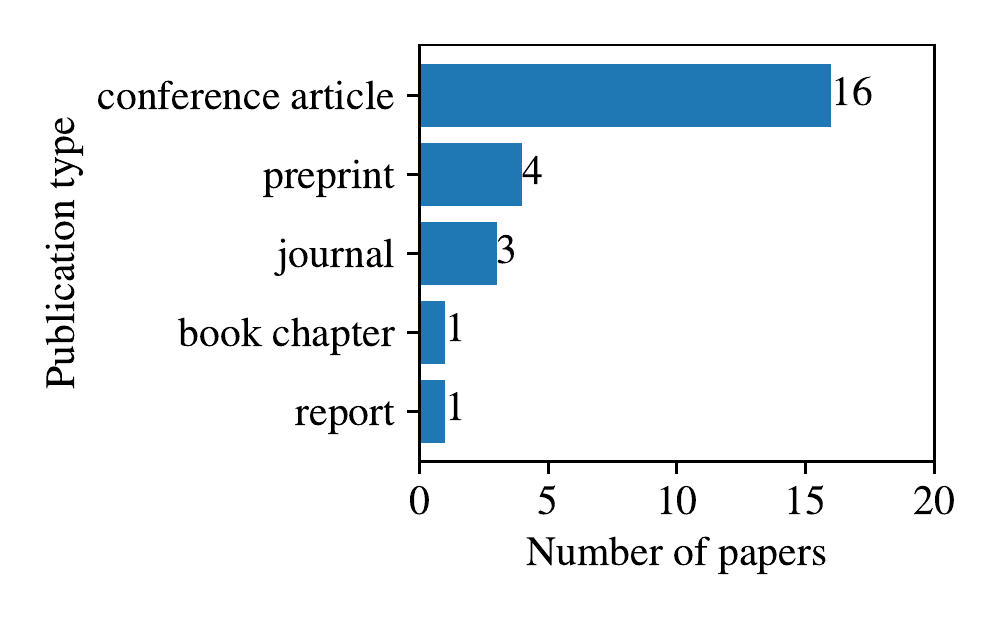}
        \caption{Types of publication selected.}
        \label{fig:publications_types}
     \end{subfigure}
    \caption{Overview of the evolution and the types of papers selected.}
    \label{fig:year_types}
\end{figure}

The next largest group of coauthors accounts for 4 papers \cite{lee_causal_2021,oh_effectively_2021,torkar_bayesian_2020,scholz_empirical_2021}. All the left 11 coauthors groups account each for a single paper. Figure \ref{fig:papers_per_year} shows the number of papers published over time and Figure \ref{fig:publications_types} provides an overview of the types of publications.

\subsection{Use cases}\label{sec:use_cases}
\begin{figure}[!htbp]
    \centering
    \includegraphics[width=0.8\textwidth]{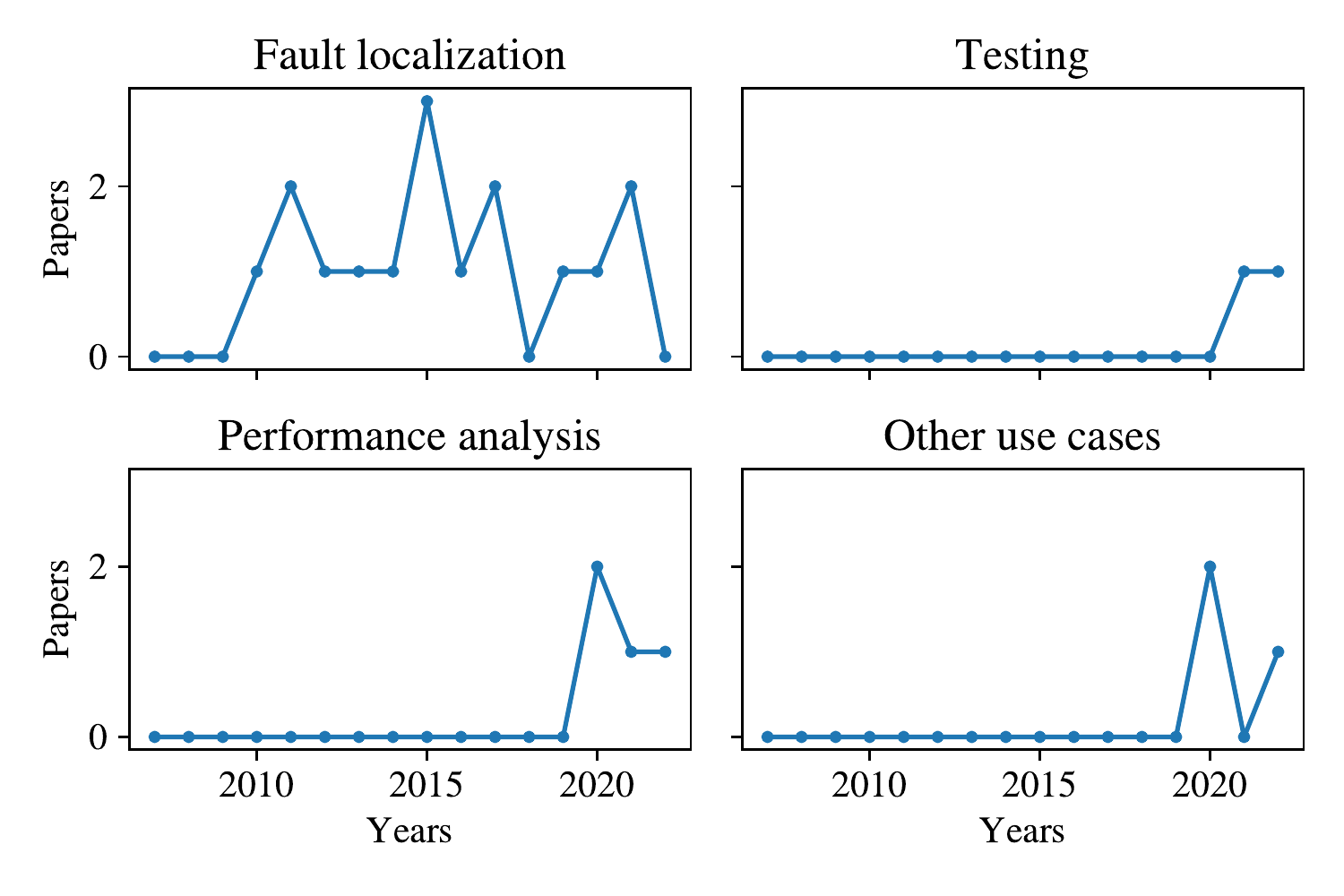}
    \caption{Number of publications over the years for each use case group.}
    \label{fig:nb_papers_per_use_case_per_year}
\end{figure}

From the \nbpapers{} selected papers, we extracted the use cases and organized them into four groups: papers focusing on fault localization (G1, see section \ref{sec:fault_loc_g1}), on testing (G2, see section \ref{sec:testing_g2}), on performance analysis (G3, see section \ref{sec:performance_g3}), and papers reporting other use cases (G4, see section \ref{sec:other_g4}). Table \ref{tab:use_cases} provides an overview of all four groups and figure \ref{fig:nb_papers_per_use_case_per_year} provides an overview of the evolution of the number of publications for each group.

\begin{table}[!htbp]
    \footnotesize
    \centering
    \begin{tabular}{|p{6.5cm}|c|p{3.5cm}|}
        \hline
        \textbf{Focus} & \textbf{Papers} & \textbf{References} \\
        \hline
        \textbf{Faults localization} & 16 & \cite{assi_acdc_2017,baah_causal_2010,baah_matching_2011,baah_mitigating_2011,bai_importance_2015,bai_numfl_2015,bai_causal_2017,feyzi_inforence_2019,gore_reducing_2012,kucuk_improving_2021,lee_causal_2021,li_causal_2014,podgurski_counterfault_2020,shu_mfl_2013,sun_properties_2016,wang_mitigating_2015}\\
        \hline
        \textbf{Testing} & 2 & \cite{clark_testing_2022,oh_effectively_2021}\\
        \hline
        \textbf{Performance analysis} & 4 & \cite{geiger_causal_2020,iqbal_unicorn_2022,scholz_empirical_2021,sruthi_pitfalls_2020}\\
        \hline
        \textbf{Other:} & 3 & \\
        Effort estimation & & \cite{torkar_bayesian_2020} \\
        Software evolution & & \cite{leidekker_causal_2020} \\
        Experience report & & \cite{issa_mattos_use_2022} \\
        \hline
    \end{tabular}
    \caption{Overview of the SE use cases for the \nbpapers{} papers selected.}
    \label{tab:use_cases}
\end{table}

\subsubsection{Fault localization (G1)}\label{sec:fault_loc_g1}
The first group includes 18 papers focused on fault localization. 
Fault localization is the task of automatically finding statements or other program elements that might potentially be the cause of a program failure.

\paragraph{Demographics}
Out of the 16 papers, 10 papers have been published by a single group of coauthors (the group of Prof. Podgurski at Case Western Reserve University, Cleveland, OH, USA, and coauthors), from 2010 until 2021 \cite{baah_causal_2010,baah_matching_2011,baah_mitigating_2011,bai_importance_2015,bai_numfl_2015,bai_causal_2017,kucuk_improving_2021,podgurski_counterfault_2020,shu_mfl_2013,sun_properties_2016}. The other 6 papers \cite{assi_acdc_2017,feyzi_inforence_2019,gore_reducing_2012,lee_causal_2021,li_causal_2014,wang_mitigating_2015} were published between 2012 and 2021, by distinct coauthors groups.

Almost all papers (14 out of 16) directly extend the work from Baah, Podgurski, and Harrold \cite{baah_causal_2010,baah_matching_2011,baah_mitigating_2011}.

\cite{lee_causal_2021} proposes causal program dependence analysis (where causal inference is used to model the strength of program dependence relations) and applies it to fault localization. The work from Baah et al. \cite{baah_causal_2010,baah_matching_2011,baah_mitigating_2011} and Gore and Reynolds \cite{gore_reducing_2012} are cited as related work.

Only one paper \cite{li_causal_2014} in this category does not cite any of the other related work from this group. It focuses on fault localization in composite services.

\paragraph{Application of SCI}
All the papers except for \cite{li_causal_2014} focus the causal effect a given source code element might have on a failure. In \cite{li_causal_2014}, the focus lies on inputs/outputs relationships in composite services.

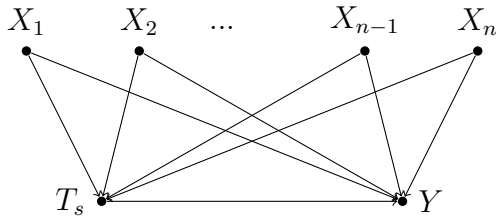
\begin{figure}[!htbp]
    \centering
        \begin{tikzpicture}
            \node (x1) at (-2,2) [circle,fill,inner sep=1.25pt,label=above:$X_1$] {};
            \node (x2) at (-0.5,2) [circle,fill,inner sep=1.25pt,label=above:$X_2$] {};
            \node (dots) at (0.6,2) [label=above:$...$] {};
            \node (xn1) at (2.5,2) [circle,fill,inner sep=1.25pt,label=above:$X_{n-1}$] {};
            \node (xn) at (4,2) [circle,fill,inner sep=1.25pt,label=above:$X_n$] {};
            \node (t) at (-1,0) [circle,fill,inner sep=1.25pt,label=left:$T_s$] {};
            \node (y) at (3,0) [circle,fill,inner sep=1.25pt,label=right:$Y$] {};
            \path[->] (t) edge (y);
            \path[->] (x1) edge (t);
            \path[->] (x1) edge (y);
            \path[->] (x2) edge (t);
            \path[->] (x2) edge (y);
            \path[->] (xn1) edge (t);
            \path[->] (xn1) edge (y);
            \path[->] (xn) edge (t);
            \path[->] (xn) edge (y);
       \end{tikzpicture}
    \caption{Illustration of the types of graphical models used by Baah et al. \cite{baah_causal_2010,baah_matching_2011,baah_mitigating_2011}. The graphical causal model contains three types of nodes: the treatment $T_s$, the outcome $Y$ and confounders $X_i$ (nodes having an effect on both $Y$ and $T_s$). All variables are binary. The outcome variable $Y$ corresponds to the outcome of a given test (1 if it fails and is 0 otherwise); the treatment variable $T_s$ corresponds to the fact that a given statement $s$ was covered by the test or not; and finally the confounders are statements that both have an impact on $s$ (through control dependence), the corresponding variable is whether the confounding statements have been covered by the test. Identification is done using the back-door criterion and measuring the causal effect of a treatment $S$ on whether a test fails or pass is computed after adjusting for confounders.}
    \label{fig:model_baah}
\end{figure}

All papers except \cite{li_causal_2014,lee_causal_2021} obtain a graphical causal models from analysis of the program structure (from control dependency or data dependency graph). Identification is done using the application of the backdoor criterion in order to reduce confounding bias. Estimation is done either through the application of linear regression, random forest, or matching techniques. Figure \ref{fig:model_baah} gives more details about the type of graphical causal model used in this work.

The work from Lee et al. \cite{lee_causal_2021} focuses first on causal program dependency analysis (CPDA). The graphical causal model represents how likely a change to the value of a program element $s_i$ is going to cause a change to the value of another element $s_j$. This model is obtained from both the analysis of the source code and interventions (mutations). Authors apply this modeling technique to the problem of fault localization. The working hypothesis is that a faulty program element will have more impact on the output element in failing executions than in passing executions. Identification is done using the backdoor criterion and adjusting for confounders. Estimation is done but no explicit causal estimation method is mentioned.

As stated above, the work from Li et al. \cite{li_causal_2014} differs from the others in this group. It focuses on fault localization not in a given program but in service composition (i.e., in a set of interacting programs). The graphical causal model is obtained from the analysis and transformation of a service dependency graph: a graph representing services as nodes and service dependencies as links. Identification is done using the backdoor criterion. Estimation is also done but no explicit causal estimation method is mentioned.

\paragraph{Evaluation methods and data sets}
Apart from \cite{li_causal_2014}, which performs a proof of concept relying upon simulated data, all other papers rely upon data sets containing programs, bugs (ground truth) and their corresponding tests suites that have been either collected for the purpose of the study or are publicly available (e.g., Defect4J, Siemens Suite, SIR). As these data sets contain a ground truth (past found bugs and the corresponding fault program elements), the papers are all performing benchmarks and compare different solutions or different versions of a solution with one another.

\subsubsection{Testing (G2)}\label{sec:testing_g2}
The second group includes two papers \cite{clark_testing_2022,oh_effectively_2021} focusing on testing. Software testing involves dynamic verification of the behaviour of a program against expected behaviour on a finite set of test cases. As such, fault localization (see section \ref{sec:fault_loc_g1}) can be considered as a subdomain of testing. The 2 papers reviewed apply causal inference for different testing methods: metamorphic testing and sensitivity analysis for \cite{clark_testing_2022}, mutation testing for \cite{oh_effectively_2021}.

\paragraph{Demographics}
The work from Oh et al. \cite{oh_effectively_2021} is about CPDA introduced in \cite{lee_causal_2021}. Two of the authors: Seongmin Lee and Shin Yoo are also coauthors of \cite{lee_causal_2021}.
The work from Clark et al. \cite{clark_testing_2022} refers to many papers reviewed in the previous section. To the best of our knowledge this is their first paper reporting on the use of SCI for testing. 

\paragraph{Application of SCI}
In the paper \cite{oh_effectively_2021}, the focus lies on mutation testing. In mutation testing, small deleterious modifications are applied to program elements (called mutations), aiming at ensuring that a software test suite is capable of detecting a maximum of mutations. Strategies for choosing program elements to mutate are based upon their causal effect computed by the CPDA approach from \cite{lee_causal_2021}. Causal identification and estimation are the same as in \cite{lee_causal_2021}.

In \cite{clark_testing_2022}, the focus lies on testing scientific simulation software, or more precisely on testing the causal effects represented in the model being simulated. A graphical causal model is created in a semi-automated way (manual pruning by experts from the simulated system is required). Nodes are inputs and outputs of the simulation program. Identification is done using the backdoor criterion. The authors use regression and causal forest estimators as estimation methods.

\paragraph{Evaluation methods and data sets}
In comparison to the previous use case, the papers deal with situations in which the ground truth is at best uncertain and at worst unknown. In \cite{clark_testing_2022}, the authors provide a proof of concept using data being generated from three scientific simulations. In \cite{oh_effectively_2021}, the authors are comparing three heuristics for generating mutants using data from the SIR data set mentioned in section \ref{sec:fault_loc_g1}.

\subsubsection{Performance modeling (G3)}\label{sec:performance_g3}
The third group includes four papers \cite{geiger_causal_2020,iqbal_unicorn_2022,scholz_empirical_2021,sruthi_pitfalls_2020} focusing on performance analysis. In \cite{geiger_causal_2020,iqbal_unicorn_2022,sruthi_pitfalls_2020}, the focus lies on performance debugging where the general goal is to understand which component of a system contributes to the measured performance and potentially to control for automated-decisions that impacts performance. \cite{geiger_causal_2020} targets cloud computing, \cite{iqbal_unicorn_2022} highly configurable systems, and \cite{sruthi_pitfalls_2020} video streaming. The work from Scholz et al. \cite{scholz_empirical_2021} is different as it aims at analysing the predictive performance of different faults prediction algorithms.
 
\paragraph{Demographics}
All papers in this category are from different research groups. Only two authors, Richard Torkar and Robert Feldt, are coauthors of other papers, which we included and reviewed: \cite{torkar_bayesian_2020} (see section \ref{sec:other_g4}) and \cite{lee_causal_2021} (see section \ref{sec:fault_loc_g1}).

The three papers \cite{geiger_causal_2020,iqbal_unicorn_2022,sruthi_pitfalls_2020} cite other works that we collected, either from C1 (causal structure discovery), from C2 (counterfactual analysis), or from C4 (causal inference) \cite{baah_causal_2010,feyzi_inforence_2019}.

The paper \cite{scholz_empirical_2021} seem to belong to a somewhat adjacent research community and does not cite any of the papers we collected.

\paragraph{Application of SCI}
Geiger et al. \cite{geiger_causal_2020} aim at addressing two problems: 1) optimizing automated-decisions (that are made during the operation of a cloud server) that can impact performance (authors called it the control problem); 2) understanding which component of a system contributes to what extent to the measured performance (authors called it the performance debugging problem). Both problems are formally defined within the language of counterfactuals and the authors propose a four-step method  to solve both problems from getting a causal graph (here the authors stay relatively generic and propose to use a mix of randomized interventional experiments, observational data, and expert knowledge and refers to other related work), until estimating the causal effects. In their application examples, nodes in the causal graph represent system and network metrics.

\cite{iqbal_unicorn_2022} focus on the problem of understanding and estimating the impact of configuration options on performance. The graphical causal model is obtained via the use of a causal structure discover algorithm (FCI). Nodes in the graph represent configuration options and system and network performance metrics. Identification is done using the do-calculus (i.e., not limited to the backdoor criterion). Estimation methods are not directly mentioned but the paper refers to the following libraries ananke\footnote{\url{https://ananke.readthedocs.io/en/latest/}} and causality\footnote{\url{https://github.com/akelleh/causality}} for estimating the causal effects.

\cite{sruthi_pitfalls_2020} focuses on analysing the performance of video streaming. The causal graph is obtained manually, and nodes represent network metrics. Identification is done using the backdoor criterion. The paper does not directly mention any estimation method.

Scholz and Torkar \cite{scholz_empirical_2021} aim at comparing the performances of different fault prediction algorithms. They use causal inference for the comparison but not directly for the fault localization and prediction (as in section \ref{sec:fault_loc_g1}). They generated the graphical causal model manually. The nodes represent the fault prediction algorithm to be compared (as treatment), evaluation metrics (as outcome), and project, language, LOC, and fix count. They applied the rules of do-calculus for the identification step. For the estimation, they used linear regression.

\paragraph{Evaluation methods and data sets}
In this group all papers rely upon a ground truth that is either being simulated (as in \cite{geiger_causal_2020,sruthi_pitfalls_2020}) or collected for the purpose of the study (as in \cite{iqbal_unicorn_2022,scholz_empirical_2021}).

\subsubsection{Other use cases (G4)}\label{sec:other_g4}
The last group includes three papers, where the use cases were different from the other three groups. \cite{torkar_bayesian_2020} applies causal inference methods to the problem of effort estimation. \cite{leidekker_causal_2020} reports on an early work on how to use SCI methods for building and evaluating software evolution theories. \cite{issa_mattos_use_2022} is a short paper reporting on the usage of graphical causal models in the automotive industry.

\paragraph{Demographics}
In \cite{torkar_bayesian_2020}, one of the coauthor also appears in another paper of this study \cite{scholz_empirical_2021} (see section \ref{sec:performance_g3}). Both authors in \cite{issa_mattos_use_2022}, David Issa Mattos and Liu Yuchu are also coauthors of two papers, which we classified as C3 (doing causal inference without relying upon graphical causal models) \cite{liu_bayesian_2021,liu_bayesian_2022}. The work in \cite{leidekker_causal_2020} does not cross-cite any other paper in our study.

\paragraph{Application of SCI}
\cite{torkar_bayesian_2020} uses a graphical causal model (obtained manually), do-calculus for identification, and linear regression for estimation. The nodes in the graph represent code and project metrics taken from the International Software Benchmarking Standards Group (ISBSG) data set\footnote{\label{isbsg}\url{isbsg.org}}.

\cite{issa_mattos_use_2022} presents how graphical causal model are used at Volvo Cars. These graphs are obtained manually from domain experts. The paper does not mention how identification or estimation is done.

\cite{leidekker_causal_2020} reproduces the modeling assumptions of two existing studies with the help of a causal diagram (obtained manually). Unfortunately, the document does not specify whether and how the identification and estimation was done.

\paragraph{Evaluation and data sets}
\cite{torkar_bayesian_2020} is using data from the International Software Benchmarking Standards Group (ISBSG). The other two papers do not perform any evaluation.

\subsection{Libraries}\label{sec:libraries}
We extracted the names and URL of the libraries used in the selected papers. When the libraries were not mentioned but the paper provided an implementation repository, we analysed the repository dependencies. Our focus lies on libraries specific to SCI. We excluded libraries to perform regression tasks (like Ranger\footnote{ \url{https://cran.r-project.org/web/packages/ranger/index.html}} or Glmnet\footnote{\url{https://cran.r-project.org/web/packages/glmnet/index.html}}), 
Bayesian probabilistic inference (like Pyro\footnote{\url{https://github.com/pyro-ppl/pyro}} or Brms \footnote{\url{https://cran.r-project.org/web/packages/brms/index.html}}),
SEM (like Semopy\footnote{\url{https://semopy.com/}}),
and libraries to visualize or to manipulate graphs (like causalgraphicalmodels \footnote{\url{https://github.com/ijmbarr/causalgraphicalmodels}}). Table \ref{tab:libraries} provides an overview of the libraries mentioned in the selected papers.

The first thing we noticed, is that almost two thirds of the papers either did not mention any library (ten papers out of \nbpapers{}) or only mentioned using the R language (10 papers out of \nbpapers{}). The last five papers mentioned six distinct libraries related to SCI.   

\begin{table}[!htbp]
    \centering
    \footnotesize
    \begin{tabular}{|ll|c|}
        \hline
        \textbf{Library name} & \textbf{URL} & \textbf{Nb.}\\
        \hline
        \multicolumn{2}{|l|}{not mentioned}  & 10 \\
        \hline
        \multicolumn{2}{|l|}{only mention using R} & 10 \\
        \hline
        dagitty (R) & \url{https://cran.r-project.org/web/packages/dagitty/index.html} & 2 \\
        \hline
        econml (Python) & \url{https://github.com/microsoft/econml} & 1 \\
        \hline
        matching (R) & \url{https://cran.r-project.org/web/packages/Matching/index.html} & 1 \\
        \hline
        matchit (R) & \url{https://cran.r-project.org/web/packages/MatchIt/index.html} & 1 \\
        \hline
        ananke (Python) & \url{https://gitlab.com/causal/ananke} & 1 \\
        \hline
        causality (Python) & \url{https://github.com/akelleh/causality} & 1 \\
        \hline
    \end{tabular}
    \caption{Causal inference libraries used. The last column corresponds to the number of papers mentioning a given library.}
    \label{tab:libraries}
\end{table}

\section{Answers to our research questions}\label{sec:answers}

\subsection{RQ1: use cases}
In section \ref{sec:use_cases}, we analysed the different SE use cases in which SCI had found an application. 
We clustered the use cases into four groups: fault localization (16 papers), testing (two papers), performance analysis (four papers) and  other use cases (three papers). Most of the papers, which we collected were oriented towards the broader task of improving software quality.

\subsection{RQ2: structure of the research community}
Based upon our results (see section \ref{sec:demographics} and figure \ref{fig:nb_papers_per_use_case_per_year}), we can say that the application of causal inference in SE is relatively young. The community of researchers applying SCI in SE seems to be fragmented. We could only find one group of researchers (the group of prof. Podgurski and coauthors at Case Western Reserve University, Cleveland, OH, USA) that applied SCI over more than ten years. As already noted in section \ref{sec:analysis}, there are not so many cross-references. Only one paper \cite{clark_testing_2022} provided an extensive state of the art (although focused on testing).

\subsection{RQ3: causal methods}
In section \ref{sec:def}, we provide a simplified SCI process with three main steps: modeling, identification and estimation.

\subsubsection{Modeling}
Concerning the modeling phase (building a graphical causal model), five papers used a manual approach  \cite{sruthi_pitfalls_2020,scholz_empirical_2021,torkar_bayesian_2020,leidekker_causal_2020,issa_mattos_use_2022}. One paper used an existing causal structure discovery algorithm (FCI) \cite{iqbal_unicorn_2022}. 20 papers have developed specific extraction methods for generating the graphical causal model in an automated way. One paper \cite{clark_testing_2022} used both an automated and an expert based (manual) approach for generating the graphical causal model.

\subsubsection{Identification and Estimation}

Almost all studied papers use the Backdoor criterion for identification (only two papers mention applying the rules of do-calculus without mentioning specific method like the Backdoor or the Frontdoor criteria). Estimation is then done, either through classical statistical estimation methods (e.g., linear regression or random forest) or matching techniques. Table \ref{tab:est_dag} provides the details of the corresponding estimation methods and papers.

\begin{table}[!htbp]
    \footnotesize
    \centering
    \begin{tabular}{|l|c|l|}
        \hline
        \textbf{Estimation method} & \textbf{Nb.} & \textbf{Ref.} \\
        \hline
        Linear regression & 9 & \cite{assi_acdc_2017,baah_causal_2010,bai_importance_2015,gore_reducing_2012,scholz_empirical_2021,shu_mfl_2013,sun_properties_2016,torkar_bayesian_2020,wang_mitigating_2015} \\
        \hline
        Random forest & 2 & \cite{kucuk_improving_2021,podgurski_counterfault_2020} \\
        \hline
        Propensity score matching & 3 & \cite{bai_numfl_2015,bai_causal_2017,feyzi_inforence_2019}\\
        \hline
        Exact matching & 2 & \cite{baah_matching_2011,baah_mitigating_2011} \\
        \hline
        Done but not explicit & 7 & \cite{lee_causal_2021,li_causal_2014,sruthi_pitfalls_2020,iqbal_unicorn_2022,geiger_causal_2020,oh_effectively_2021,clark_testing_2022} \\
        \hline
        Not done & 2 & \cite{leidekker_causal_2020,issa_mattos_use_2022}\\
        \hline
    \end{tabular}
    \caption{Estimation methods for papers relying upon an explicit graphical causal model.}
    \label{tab:est_dag}
\end{table}

\subsection{RQ4 and RQ5: Data sets, libraries and evaluation}
The results show that causal inference provides improvement, usually, through mitigating confounding bias (as illustrated by the benchmarks in \cite{kucuk_improving_2021,oh_effectively_2021}). Most of the reviewed papers use specific data sets that contains a ground truth for the specific use case at hand.

In section \ref{sec:libraries}, we provided an overview of the libraries used. Many papers (20 out of \nbpapers{}) do not mention any specific SCI library. Out of these, 10 simply mention using R or using packages for regression (like linear regression or random forest). This can be explained by the fact that, for these papers, identification boils down to adjusting for the parents of the treatment variable under study. However, it must be noted that when the number of adjusted confounders increases, classical regression reaches its limits and specific estimation methods such as metalearners like X-learner might prove to perform better \cite{kuenzel_metalearners_2019}. 

The distribution among the selected papers shows no particular preferred library and an equal mix between Python and R packages. 

We have not been able to find any paper mentioning a specific benchmark data set dedicated to SCI in SE (as one could find in \cite{yao2021survey}).

\section{Discussions}\label{sec:discussion}
\subsection{Application scope}
Not all questions are aimed to be answered with a single tool. In his lecture and subsequent paper "Why Model?" \cite{epstein2008whymodel}, Epstein lists 17 different types of scientific questions (he called them modeling purposes). In order to situate the type of questions that can be answered with SCI methods, we can refer to Hernan et al. \cite{hernan2019secondchance} and Pearl's ladder of causation \cite{pearl2019book}. Both frameworks highlight the difference between tasks that do not require the use of a causal model such as description: \textit{"what is $Y$?"} and prediction \textit{"if I see X, what can I tell about Y?"}, and tasks that do require one such as intervention: \textit{"what if I do $X$, what is the effect on $Y$?"} and counterfactuals \textit{"what if, in the same circumstances, instead of having done $X$, I had done $X'$?"}. There is a hierarchical structure between these tasks, and Pearl showed that the formal tools used to answer a prediction question are not sufficient to answer an intervention question, and that the formal tools used to answer an intervention question are not sufficient to answer a counterfactual question \cite{pearl2019book}.  

Some use cases in SE can be easily mapped to one of the four aforementioned tasks (description, prediction, intervention, and counterfactuals). For example, mutation testing is a typical example of a counterfactual analysis, as it involves testing a program under the same circumstances, but for a statement that has been mutated. Other questions like fault localization are not so obvious. At first, fault localization might seem like a prediction problem: given a program, predict which statement is more likely to induce a fault (i.e., its suspiciousness score). The papers reviewed in section \ref{sec:fault_loc_g1} used the estimated causal effect of a statement on the failure of a given test to compute its suspiciousness score. There is no absolute correspondence between the SE use cases and the type of task (i.e. description, prediction, intervention and counterfactuals). Each task can be useful within a single use case. Each of these tasks simply has its limits and it is necessary to know how to recognize them and use the corresponding methods in the appropriate context.

\subsection{Obtaining a causal graphical model}
To take full advantage of Pearl's SCI framework, a graphical causal model is required. This model can be obtained manually from domain experts or from previous experiments, it can be extracted from the software structure (e.g. from static code analysis, see section \ref{sec:fault_loc_g1}), or it can be learned from data (e.g. by mining software repositories or analysing logs, see also the papers classified as C1, see table \ref{tab:category_c1}). The main purpose of such a graph is to reason about the presence of potential spurious correlations that may bias the analysis. It may not always be possible to obtain a detailed graph, as some influencing factors may not be directly measurable (such as "developer experience"), or in a high-dimensional setting it may be unclear what the causal structure looks like. In these cases, causal structure learning algorithms can help extract a causal structure from the data. It should be noted that there are different methods that rely on different assumptions, maybe the most important being whether we believe that all relevant variables have been measured or whether there are some unmeasured variables left. This is, for example, a major difference between two classical causal structure discovery algorithms: the PC algorithm, which assumes that there are no unobserved confounders, and FCI, which assumes the presence of unobserved confounders \cite{glymour2019review}.
Even if the underlying causal graph is only partially known, it is already helpful to reason about potential bias and which estimation method might work (see section \ref{ref:other_sci_methods}). As always with analysis methods, the devil is in the detail, and several papers, such as \cite{guo_2020_propensity,huenermund_doubleml_2021}, remind us to be cautious about applying these methods without a proper understanding of their underlying assumptions (for an easy-to-read introduction to the pros and cons of the various SCI methods, see part 2 of \cite{huntington2021effect}).

\subsection{Reducing spurious correlation bias}
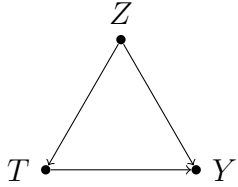
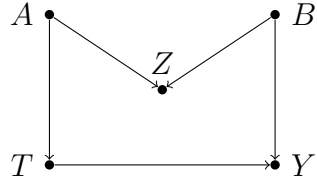
\begin{figure}
    \centering
    \begin{subfigure}[b]{0.45\textwidth}
         \centering
         \begin{tikzpicture}
            \node (z) at (60:2) [circle,fill,inner sep=1.25pt,label=above:$Z$] {};
            \node (t) at (0:0) [circle,fill,inner sep=1.25pt,label=left:$T$] {};
            \node (y) at (0:2) [circle,fill,inner sep=1.25pt,label=right:$Y$] {};
            \path[->] (t) edge (y);
            \path[->] (z) edge (t);
            \path[->] (z) edge (y);
       \end{tikzpicture}
         \caption{Confounder bias. $Z$ is a confounder and has an effect on both the treatment $T$ and the outcome $Y$. Not controlling for $Z$ will let spurious correlation flows along the path $\{T \leftarrow Z \rightarrow Y\}$.}
         \label{fig:confounding}
     \end{subfigure}
     \hfill
     \begin{subfigure}[b]{0.45\textwidth}
         \centering
         \begin{tikzpicture}
            \node (a) at (0,2) [circle,fill,inner sep=1.25pt,label=left:$A$] {};
            \node (b) at (3,2) [circle,fill,inner sep=1.25pt,label=right:$B$] {};
            \node (z) at (1.5,1) [circle,fill,inner sep=1.25pt,label=above:$Z$] {};
            \node (t) at (0,0) [circle,fill,inner sep=1.25pt,label=left:$T$] {};
            \node (y) at (3,0) [circle,fill,inner sep=1.25pt,label=right:$Y$] {};
            \path[->] (t) edge (y);
            \path[->] (a) edge (t);
            \path[->] (b) edge (y);
            \path[->] (a) edge (z);
            \path[->] (b) edge (z);
       \end{tikzpicture}
         \caption{M-bias: controlling for the collider $Z$ will open a spurious correlation path $\{T\leftarrow A \rightarrow Z \leftarrow B \rightarrow Y\}$ that is otherwise blocked.}
         \label{fig:mbias}
     \end{subfigure}
    \caption{Two typical examples of how bias can happen (see \cite{cinelli2022crashcourse} for other examples).}
    \label{fig:bias}
\end{figure}

An argument made by many selected papers for the use of SCI methods over more traditional statistical inference techniques is the need to recognize and manage spurious correlation biases, for example confounder bias. A confounder is a variable that both has an impact on the treatment and the outcome of interest (see Figure \ref{fig:confounding}). Note that confounder bias is not the only bias that statistical causal methods can mitigate (another one is called M-bias, see Figure \ref{fig:mbias}), but to mitigate them, one needs a graphical causal model. The authors in \cite{cinelli2022crashcourse} provide an analysis of typical bias and their corresponding graphical structure. 

It is difficult to generalize the improvement provided by the application of SCI for all the use cases reviewed. Only in the case of fault localization (G1, see section \ref{sec:fault_loc_g1}), do we have enough evaluation results from independent researchers group. Out of 16 papers, 11 used a similar evaluation approach. They first compute the effectiveness of the proposed fault localization methods based on the EXAM score (also called cost). "EXAM score is the percentage of program statements a developer must examine, in non-increasing order of their suspiciousness scores, before finding the fault" \cite{kucuk_improving_2021}. The evaluation is done by comparing several fault localization approaches on benchmark data sets (see the details in table \ref{tab:eval_fault_loc}). All these papers have shown that on average SCI methods improve fault localization in comparison to other more classical approaches. 

\begin{table}
    \centering
    \footnotesize
    \begin{tabular}{|c|p{5cm}|p{8.5cm}|}
        \hline
        \textbf{Ref.} & \textbf{Compared methods} & \textbf{Test programs}\\
        \hline
        \cite{baah_causal_2010} & Tarantula, Ochiai, F1-measure, Importance(p) & Print-Tokens, Print-Tokens2, Replace, Schedule, Schedule2, Tcas, Tot-Info, Sed, Space \\
        \hline
        \cite{baah_mitigating_2011} & Tarantula, Ochiai & Cal, Col, Comm, Spline, Tr, Uniq, Print-Tokens, Print-Tokens2, Replace, Schedule, Schedule2, Tcas, Tot-Info, Sed, Space \\
        \hline
        \cite{gore_reducing_2012} & Tarantula, Ochiai, F1-measure & Tcas, Tot-Info, Schedule, Schedule2, Print-Tokens, Print-Tokens2, Replace, Sed, Space, Bc, Gzip \\
        \hline
        \cite{shu_mfl_2013} & Tarantula, Ochiai, PFIC, F1-measure &	ROME, Xerces2, XStream, HRL \\
        \hline
        \cite{wang_mitigating_2015} & Tarantula, Ochiai, Naish1, Wong1, CBI and SOBER & Jtcas, Tot-Info, Schedule1, Schedule2, Print-Tokens1, Print-Tokens2, NanoXML 1, NanoXML 2, NanoXML 3, NanoXML 5, Siena 1, Siena 2, Siena 5, Siena7 \\
        \hline
        \cite{bai_numfl_2015,bai_causal_2017} & Ochiai, DStar, SOBER, Exploratory Software Predictor (ESP) & Apache\_EigenDecompose, Apache\_DScompiler, Apache\_BigMatrix, Apache\_Rotation3D, Ojaljo\_SchurDecompose, Jama\_MatrixDecompose, SciMark\_LU, SciMart\_FFT, Apache\_SymmLQ,  Apache\_SplineInterpolator, Apache\_SimpleRegress, Apache\_SchurTransformer, Apache\_MillerUpdatRegress, Apache\_HarmonicFitter, Apache\_FastSine, Apache\_FastCosine \\
        \hline
        \cite{sun_properties_2016} & Symmetric Klosgen (SK), relative Ochiai (RO), relative F1-measure (RF1), relative recall (RR), enhanced Tarantula (ET) & Daikon 4.6.4, Eventbus 1.4, Jaxen 1.1.5, Jster 1.37b, JExel 1.0.0b13, Jparsec 2.0, AC codec 1.3, AC Lang 3.0, EclipseDraw2d, HTML Parser 1.6 \\
        \hline
        \cite{feyzi_inforence_2019} & D-star, Crosstab, H3b-H3c, Ochiai, OOP, Relief, Gen-Entropy, Baah’s \cite{baah_causal_2010,baah_mitigating_2011}, GP02, GP03, GP19 & Schedule, Schedule2, Print-Tokens, Print-Tokens2, Replace, Tot-Info, Tcas, Total, Gzip, Grep, Sed, Space, Make, Bash \\
        \hline
        \cite{podgurski_counterfault_2020} & NUMFLQRM, NUMFLDLRM, Elastic Predicates (ESP), Klosgen, F1-measure, Ochiai, D-Star & Chart, Math, Time, Lang, FastSineTransformer, SplineInterpolator, FastCosineTransformer, SchurTransformer, HarmonicFitter, SimpleRegression, MillerUpdateRegression, Rotation and Vector3D, EigenvalueDecomposition, DSCompiler \\
        \hline
        \cite{kucuk_improving_2021} & NUMFLQRM, NUMFLDLRM, Elastic Predicates (ESP), Baah's \cite{baah_causal_2010}, Predicate Switching, Ochiai, D-Star & Chart, Cli, Csv, Math, Time, Lang, Closure, Mockito, Codec, JxPath, Gson, Collections, Compress, Jsoup, JacksonCore, JacksonXml, JacksonDatabind\\
        \hline
    \end{tabular}
    \caption{Details of the evaluations done in papers focusing on fault localization. Here only comparable papers from G1 are shown.}
    \label{tab:eval_fault_loc}
\end{table}
 
Relatively few papers have tried to analyse why SCI methods improve fault localization. First, Baah, Podgurski and Harrold \cite{baah_causal_2010} have shown that SCI can reduce confounding bias in fault localization scores. In their following work \cite{baah_mitigating_2011}, the authors showed that using control dependence graphs does not capture all sources of confounders and that data dependencies need to be taken into account. A following work by Bai, Sun, and Podgurski analysed the impact of violating an important precondition for causal inference, called positivity \cite{bai_importance_2015}. Finally, in their most recent work, Küçük, Henderson, and Podgurski analysed the effect of covariate balance and imbalance on the performance fault localization methods. These papers provide us with an initial understanding of the benefits and limitations of SCI methods for fault location. Similar research will be needed to understand their potential and limitations in the context of other SE use cases.

\subsection{Side effect: make some hypotheses explicit}
An important aspect to mention is that, in order to make use of SCI methods, it is necessary to make assumptions explicit: both about the aspects that are part of the system under study and about the assumed structure of the causal model. This point was also mentioned in two of our selected papers: \cite{torkar_bayesian_2020,issa_mattos_use_2022} and is a recuring theme in empirical research.

\subsection{Implications for SE research}
SCI methods belong to the toolkit of empirical research and are well suited for dealing with observational data. We believe, like \cite{deoliveiraneto2019evolution}, that in the future such methods will become more mainstream in SE research. However, it appears that these methods have been either relatively unknown until now or, if known, deemed unnecessary or too complicated to apply. One assumption is that the learning curve is relatively steep. It is only recently that books for a wider audience have been published (like \cite{pearl2019book,hernan2020whatif,huntington2021effect,cunningham_mixtape_2021}). Note that, in the data science and machine learning community several initiatives exist both from academia and industry to push forward the research and the application of SCI methods\footnote{See for instance \url{https://www.causalscience.org/} or \url{https://www.pywhy.org/}}. Chances are that these initiatives will spread to the SE community (if they haven't already) and that the learning curve will get shallower.

The main obstacle that we see for the application of SCI methods is to obtain a causal graphical model and the corresponding observational data. Even if for some use cases it is possible to extract a causal graph automatically (like in \cite{lee_causal_2021,kucuk_improving_2021,iqbal_unicorn_2022}), very few papers have actually made their code available. This hinders both the reproducibility and learning of these methods. We also have not been able to find any data set specific to SE consisting of causal graphs and the corresponding observational data. For assessing SCI methods in SE, as well as for teaching them, it would be beneficial to have dedicated benchmark data sets where one could find ground truth data and the corresponding causal graphical model(s). A recommendation would be to gather causal graphs from the literature, starting with the application of Bayesian network in SE (see section \ref{sec:related_work}) or papers in C1 (see table \ref{tab:category_c1}).

After getting a causal graphical model, it is necessary to ensure that the assumptions behind SCI are met. We have encountered only few papers that discussed the applicability conditions of these methods. A further recommendation would be to gather both experience from researchers having applied SCI in SE in order to develop methodological guidelines and examples for applying SCI in SE.

Finally, researchers in the domain of causal inference has provided specific estimation methods (such as T-learner, X-learner, etc.) that have advantages over classical regression and matching techniques. Since not all of the selected papers made use of these particular techniques, it is not yet known what the effect of their use will be. 

\subsection{Threats to validity}
\paragraph{Research design validity}
In this paper, we aimed at understanding how SCI methods have been applied in the field of SE. We conducted a mapping study, focusing on published literature (both peer-reviewed and non-peer-reviewed). This research design has its limits. 

First, it is possible that researcher have applied causal inference methods but did not publish about it. It is also quite possible that researchers in the field of SE know about Pearl's framework for SCI, but did not consider to use it. A literature review process will never get this kind of information. Only interviews will. 

Second, we have explicitly left out student work like master or PhD theses. The reason is that, although, theses are potentially interesting sources for understanding how a given method is applied, not all theses are freely available. During the search process, the only theses, we came across were from authors of papers already selected, and therefore we decided to leave them out.

Finally, we decided to include non-peer-reviewed work (in total 1 technical report and 4 preprints available on \url{https://arxiv.org}). Although, they were not subjected to a peer review process, the reason we kept them is that our interest is more in the methodology applied than in the final results.

\paragraph{Publication selection validity}

It needs to be said, that it was relatively difficult to find relevant papers (as shown in \ref{sec:methodogy}).
First, SE is a broad field, and it is possible that we might have missed some SE related aspects in our search process. Moreover, the boundaries between SE and artificial intelligence (AI) are somewhat blurred. For example, there are now many papers using causal inference for fairness (like \cite{sun_causality_2022,salimi_database_2020,salimi_interventional_2019}). Although, we found these papers during the selection process, we decided to leave them out considering them closer to AI than SE. To make matters worse, there is no standard vocabulary in SE for talking about causal inference. For instance, several papers from C1 (see table \ref{tab:category_c1}) referred to causal structure discovery as "causal inference". Papers from C2 (see table \ref{tab:category_c2}) referred to counterfactual analysis as "causal analysis" or "causal testing". 

Furthermore, as our results show, the research community using SCI in SE seems to be fragmented, therefore the snowballing approach, we used to find new relevant papers also has its limits. To mitigate these risks, we used two different types of queries and applied two snowballing phases (see section \ref{sec:methodogy}). We stopped the snowballing after two rounds because we reached a saturation, i.e., further searches only resulted in articles that were already selected.

Finally, the selection has been done by a single person.

\paragraph{Data validity}
For this paper, we tried as much as possible, to extract only factual information that is not subject to interpretation. We have not considered the quality of the publications. The reason is that our focus lies on the methods applied and less on the results achieved. Considering the value of SCI methods for the different use cases, almost all papers shows an improvement compared to their own benchmark (which raises the question of publication bias). Considering the fact that half of the papers (12 out of \nbpapers{}) were published from 2020 onwards, it is yet too early to judge properly the advantages of SCI methods. Only in fault localization, where the papers span from 2010 until now, and where independent groups of researchers have applied SCI methods can we estimate the pros and cons of SCI.

\paragraph{Data synthesis and conclusions validity}
As already mentioned, this paper is the work of a single person. The main threat to validity comes from the expertise of the reviewer. To mitigate such bias, the findings were also presented to and discussed with some of the authors of the selected papers and within a working group on causal inference. None of the persons contacted could bring novel work to our attention. Obviously, this only consolidates the validity of the conclusion but does not remove completely the threats mentioned above.

\section{Conclusion}\label{sec:conclusion}
In this paper, we conducted a mapping study and selected \nbpapers{} papers applying SCI methods in the SE domain. We focused specifically on Pearl's framework that rely upon graphical causal models. These methods are interesting when targeting problems requiring estimating causal effects from observational data. Our analysis shows that these techniques are not yet mainstream in SE research. By providing this overview, we hope to promote awareness of these methods, lay the groundwork for discussions on where and when to apply them, and eventually increase the number of applications of these methods in SE.

\section*{Acknowledgement}
The author would like to thank Nico Cappel, Niklas Gutting, and Ilir Hulaj for their preliminary work "Anwendungen von kausaler Inferenz in Software-Engineering" done during their SE Seminar at the Technical University of Kaiserslautern; Adam Trendowicz and Andreas Jedlitschka (Fraunhofer IESE, Kaiserslautern, Germany), and Andy Podgurski (Case Western Reserve University, Cleveland, OH, USA), for the useful discussions; and Patricia Kelbert for proofreading the manuscript.

\bibliographystyle{elsarticle-num}
\bibliography{main}

\newpage
\appendix
\section{Keywords analysis details}
\begin{table}[!htbp]
 \centering
 \footnotesize
 \begin{tabular}{|p{0.35\textwidth}|p{0.35\textwidth}|p{0.1\textwidth}|p{0.1\textwidth}|}
    \hline
    SE Term (\texttt{SE TERM}) & Causal Inference Term (\texttt{CI TERM}) & Number of results & Number of relevant papers \\
    \hline
    "SE" & "causal inference" & 20 & 11 \\
    \hline
    "software development" & "causal inference" & 4 & 2 \\
    \hline
    "software test*" & "causal inference" & 13 & 8 \\
    \hline
    "software design" & "causal inference" & 5 & 3 \\
    \hline
    "software architecture" & "causal inference" & 1 & 1 \\
    \hline
    "software *bug*" & "causal inference" & 3& 2 \\
    \hline
    "software requirement" & "causal inference" & 0 & 0 \\
    \hline
    "software quality" & "causal inference" & 1 & 1 \\
    \hline
    "software construct*" & "causal inference" & 0 & 0 \\
    \hline
    "technical debt" & "causal inference" & 0 & 0 \\
    \hline
    "fault local*" & "causal inference" & 19 & 17 \\
    \hline
    "config* manag*" & "causal inference" & 0 & 0 \\
    \hline
    "software process" & "causal inference" & 1 & 0 \\
    \hline
     "dev-ops"  OR  "devops"  OR  "mlops"  OR  "ml-ops"  OR  "aiops" & "causal inference" & 1 & 1 \\
    \hline
    \hline
    "SE" & "do calculus" & 1 & 0 \\
    \hline
    "SE" & "structural causal model" & 2 & 0 \\
    \hline
    "SE" & "causal graph" & 8 & 0 \\
    \hline
    "SE" & "backdoor criteri*" OR "frontdoor criteri*" & 0 & 0 \\
    \hline
    "SE" & "potential outcomes" & 9 & 2 \\
    \hline
    "SE" & "propensity score" & 10 & 5 \\
    \hline
    "SE" & "causal identification" & 0 & 0 \\
    \hline
    "SE" & "causal estimation" & 1 & 0 \\
    \hline
    "SE" & "counterfact*" & 18 & 9 \\
    \hline
    "SE" & "mediation analysis" & 2 & 0 \\
    \hline
    "SE" & "difference in difference" & 2 & 0 \\
    \hline
    "SE" & "instrument* variable" & 4 & 0 \\
    \hline
    "SE" & "inverse probability" OR "probability weigh*" & 5 & 2 \\
    \hline
    "SE"  &  "treatment effect" & 5 & 1 \\
    \hline
 \end{tabular}
 \caption{Queries used to determine the most relevant keywords. Each query was performed in Scopus on the 2nd of August 2022, targeting only Title, Abstract and Keywords (i.e., using \texttt{TITLE-ABS-KEY ()}). Each query takes the form \texttt{TITLE-ABS-KEY((SE TERM) AND (CI TERM))}.}
 \label{tab:prep-query}
\end{table}

\section{details of the four categories c1, c2, c3, and c4}
\begin{table}[!htbp]
    \footnotesize
    \centering
    \begin{tabular}{|p{0.95\textwidth}|c|}
        \hline
        \textbf{title (alphabetical order)} & \textbf{ref.} \\
        \hline
        A quantitative causal analysis for network log data & \cite{jarry_quantitative_2021} \\
        \hline
        An influence-based approach for root cause alarm discovery in telecom networks & \cite{zhang_influence-based_2021} \\
        \hline
        Causal analysis for performance modeling of computer programs & \cite{lemeire_causal_2007} \\
        \hline
        Causal analysis of network logs with layered protocols and topology knowledge & \cite{kobayashi_causal_2019} \\
        \hline
        Causal inference techniques for microservice performance diagnosis: evaluation and guiding recommendations & \cite{wu_causal_2021} \\
        \hline
        Causal modeling, discovery, and inference for SE & \cite{kazman_causal_2017} \\
        \hline
        Causal program slicing & \cite{gore_causal_2009} \\
        \hline
        Comparative causal analysis of network log data in two large ISPs & \cite{kobayashi_comparative_2022} \\
        \hline
        Detecting causal structure on cloud application microservices using Granger causality models & \cite{wang_detecting_2021} \\
        \hline
        Discovering and utilising expert knowledge from security event logs & \cite{khan_discovering_2019}\\
        \hline
        Discovering many-to-one causality in software project risk analysis & \cite{chen_discovering_2014} \\
        \hline
        Evaluation of causal inference techniques for AIOps & \cite{arya_evaluation_2021} \\
        \hline
        Falcon: differential fault localization for SDN control plane & \cite{yu_falcondifferential_2019} \\
        \hline
        Further causal search analyses with UCC's effort estimation data & \cite{hira_further_2018} \\
        \hline
        Localization of operational faults in cloud applications by mining causal dependencies in logs using golden signals & \cite{aggarwal_localization_2021} \\
        \hline
        Microdiag: fine-grained performance diagnosis for microservice systems & \cite{wu_microdiag_2021} \\
        \hline
        Mining causality of network events in log data & \cite{kobayashi_mining_2018} \\
        \hline
        Mining causes of network events in log data with causal inference & \cite{kobayashi_mining_2017} \\
        \hline
        Mutation-based graph inference for fault localization & \cite{musco_mutation-based_2016} \\
        \hline
        Preliminary causal discovery results with software effort estimation data & \cite{hira_preliminary_2018} \\
        \hline
        Software project risk analysis using Bayesian networks with causality constraints & \cite{hu_software_2013} \\
        \hline
        Thinking inside the box: differential fault localization for SDN control plane & \cite{li_thinking_2019} \\
        \hline
    \end{tabular}
    \caption{Papers classified as C1, focusing on modeling potential causal structures but not using identification and estimation techniques as defined in section \ref{sec:def}.}
    \label{tab:category_c1}
\end{table}

\begin{table}[!htbp]
    \footnotesize
    \centering
    \begin{tabular}{|p{0.95\textwidth}|c|}
        \hline
        \textbf{title (alphabetical order)} & \textbf{ref.} \\
        \hline
        A causality analysis framework for component-based real-time systems & \cite{wang_causality_2013} \\
        \hline
        A general framework for blaming in component-based systems & \cite{gossler_general_2015} \\
        \hline
        A general trace-based framework of logical causality & \cite{gossler_general_2014} \\
        \hline
        A hybrid approach to causality analysis & \cite{wang_hybrid_2015} \\
        \hline
        Causal reasoning for safety in Hennessy Milner logic & \cite{caltais_causal_2020} \\
        \hline
        Causality analysis and fault ascription in component-based systems & \cite{gossler_causality_2020} \\
        \hline
        Causality analysis for concurrent reactive systems (extended abstract) & \cite{dimitrova_causality_2019} \\
        \hline
        Causality-guided adaptive interventional debugging & \cite{fariha_causality-guided_2020} \\
        \hline
        Code-change impact analysis using counterfactuals & \cite{peralta_code-change_2011} \\
        \hline
        Code-change impact analysis using counterfactuals: theory and implementation & \cite{peralta_code-change_2013} \\
        \hline
        Counterfactually reasoning about security & \cite{peralta_counterfactually_2011} \\
        \hline
        Explaining counterexamples using causality & \cite{beer_explaining_2012} \\
        \hline
        Fault ascription in concurrent systems & \cite{gossler_fault_2016} \\
        \hline
        From probabilistic counterexamples via causality to fault trees & \cite{kuntz_probabilistic_2011} \\
        \hline
        From verification to causality-based explications & \cite{baier_verification_2021} \\
        \hline
        Symbolic causality checking using bounded model checking & \cite{beer_symbolic_2015} \\
        \hline
    \end{tabular}
    \caption{Papers classified as C2, focusing on counterfactual analysis but not relying upon any methods defined in section \ref{sec:def}.}
    \label{tab:category_c2}
\end{table}

\begin{table}[!htbp]
    \footnotesize
    \centering
    \begin{tabular}{|p{0.95\textwidth}|c|}
        \hline
        \textbf{title (alphabetical order)} & \textbf{ref.} \\
        \hline
        Bayesian causal inference in automotive SE and online evaluation & \cite{liu_bayesian_2022} \\
        \hline
        Bayesian propensity score matching in automotive embedded SE & \cite{liu_bayesian_2021} \\
        \hline
        Do developers discover new tools on the toilet? & \cite{murphy-hill_developers_2019} \\
        \hline
        Gender differences and bias in open source: pull request acceptance of women versus men & \cite{terrell_gender_2017} \\
        \hline
        License choice and the changing structures of work in organization owned open source projects & \cite{medappa_license_2017} \\
        \hline
        On software productivity analysis with propensity score matching & \cite{tsunoda_software_2017} \\
        \hline
        PERFCE: performance debugging on databases with chaos engineering-enhanced causality analysis & \cite{ji_perfce_2022} \\
        \hline
    \end{tabular}
    \caption{Papers classified as C3, doing causal inference without relying upon a causal graphical model.}
    \label{tab:category_c3}
\end{table}

\begin{table}[!htbp]
    \footnotesize
    \centering
    \begin{tabular}{|p{0.95\textwidth}|c|}
        \hline
        \textbf{title (alphabetical order)} & \textbf{ref.} \\
        \hline
        ACDC: altering control dependence chains for automated patch generation & \cite{assi_acdc_2017} \\
        \hline
        An empirical study of Linespots: a novel past-fault algorithm & \cite{scholz_empirical_2021} \\
        \hline
        Bayesian data analysis in empirical SE: the case of missing data & \cite{torkar_bayesian_2020} \\
        \hline
        Causal inference based fault localization for numerical software with NUMFL & \cite{bai_causal_2017} \\
        \hline
        Causal inference based service dependency graph for statistical service fault localization & \cite{li_causal_2014} \\
        \hline
        Causal inference for data-driven debugging and decision making in cloud computing & \cite{geiger_causal_2020} \\
        \hline
        Causal inference for statistical fault localization & \cite{baah_causal_2010} \\
        \hline
        Causal inference for theory building in software evolution work in progress & \cite{leidekker_causal_2020} \\
        \hline
        Causal program dependence analysis & \cite{lee_causal_2021} \\
        \hline
        Counterfault: value-based fault localization by modeling and predicting counterfactual outcomes & \cite{podgurski_counterfault_2020} \\
        \hline
        Effectively sampling higher order mutants using causal effect & \cite{oh_effectively_2021} \\
        \hline
        Improving fault localization by integrating value and predicate based causal inference techniques & \cite{kucuk_improving_2021} \\
        \hline
        Inforence: effective fault localization based on information-theoretic analysis and sci & \cite{feyzi_inforence_2019} \\
        \hline
        Matching test cases for effective fault localization & \cite{baah_matching_2011} \\
        \hline
        MFL: method-level fault localization with causal inference & \cite{shu_mfl_2013} \\
        \hline
        Mitigating the confounding effects of program dependences for effective fault localization & \cite{baah_mitigating_2011} \\
        \hline
        Mitigating the dependence confounding effect for effective predicate-based statistical fault localization & \cite{wang_mitigating_2015} \\
        \hline
        NUMFL: localizing faults in numerical software using a value-based causal model & \cite{bai_numfl_2015} \\
        \hline
        On the use of causal graphical models for designing experiments in the automotive domain & \cite{issa_mattos_use_2022} \\
        \hline
        Pitfalls of data-driven networking: a case study of latent causal confounders in video streaming & \cite{sruthi_pitfalls_2020} \\
        \hline
        Properties of effective metrics for coverage-based statistical fault localization & \cite{sun_properties_2016} \\
        \hline
        Reducing confounding bias in predicate-level statistical debugging metrics & \cite{gore_reducing_2012} \\
        \hline
        Testing causality in scientific modelling software & \cite{clark_testing_2022} \\
        \hline
        The importance of being positive in causal statistical fault localization: important properties of Baah et al.'s CSFL regression model & \cite{bai_importance_2015} \\
        \hline
        Unicorn: reasoning about configurable system performance through the lens of causality & \cite{iqbal_unicorn_2022} \\
        \hline
    \end{tabular}
    \caption{Papers classified as C4, applying the three steps of causal inference: modeling, identification and estimation.}
    \label{tab:category_c4}
\end{table}

\end{document}